\documentclass[conference]{IEEEtran}
\IEEEoverridecommandlockouts
\usepackage{cite}
\usepackage{amsmath,amssymb,amsfonts}
\usepackage{algorithmic}
\usepackage{graphicx}
\usepackage{textcomp}
\usepackage{xcolor}
\usepackage{booktabs}
\usepackage{threeparttable}
\usepackage{algorithm}
\usepackage{url}
\begin{document}

\title{SAMUeL: Efficient Vocal-Conditioned Music Generation via Soft Alignment Attention and Latent Diffusion}

\author{\IEEEauthorblockN{Hei Shing Cheung}
\IEEEauthorblockA{\textit{Division of Engineering Science} \\
\textit{University of Toronto}\\
Toronto, Canada \\
hayson.cheung@mail.utoronto.ca}
\and
\IEEEauthorblockN{Boya Zhang}
\IEEEauthorblockA{\textit{Division of Engineering Science} \\
\textit{University of Toronto}\\
Toronto, Canada \\
erhuboya.zhang@mail.utoronto.ca}
\and
\IEEEauthorblockN{Jonathan H. Chan}
\IEEEauthorblockA{\textit{Innovative Cognitive Computing Center} \\
\textit{King Mongkut's University}\\
\textit{of Technology Thonburi}\\
Bangkok, Thailand \\
jonathan.chan@mail.kmutt.ac.th}
}

\maketitle

\begin{abstract}
We present SAMUeL (Soft Alignment Music U-net Latent diffusion)\footnote{Publicly available at: \url{https://github.com/HaysonC/SAMUeL-GEN}}, a lightweight latent diffusion model for vocal-conditioned musical accompaniment generation that addresses critical limitations in existing music AI systems. SAMUeL introduces a novel soft alignment attention mechanism that adaptively combines local and global temporal dependencies based on diffusion timesteps, enabling efficient capture of multi-scale musical structure. Operating in the compressed latent space of a pre-trained variational autoencoder, SAMUeL achieves a 220× parameter reduction compared to state-of-the-art systems while delivering 52× faster inference. Experimental evaluation demonstrates competitive performance with only 15M parameters, outperforming OpenAI Jukebox in production quality and content unity while maintaining reasonable musical coherence. The ultra-lightweight architecture enables real-time deployment on consumer hardware, making AI-assisted music creation accessible for interactive applications and resource-constrained environments.
\end{abstract}


\section{Introduction}

Contemporary music generation systems primarily rely on text-conditioned generation \cite{agostinelli2023musiclm, copet2023musicgen} or symbolic music representation \cite{dong2018musegan}, creating a significant gap for musicians who conceptualize music through melodic ideas rather than textual descriptions. This limitation is particularly problematic for composers who begin with vocal melodies and require appropriate instrumental accompaniments—a common workflow in music production that existing AI systems fail to address effectively.

Current approaches can be categorized into three paradigms: text-conditioned models that translate natural language into audio \cite{agostinelli2023musiclm, evans2024stable}, autoregressive systems that generate music sequentially \cite{borsos2023audiolm}, and symbolic models operating on MIDI representations \cite{dong2018musegan}. While these systems demonstrate impressive capabilities, they fail to support direct musical input conditioning. The semantic gap between musical concepts and textual descriptions, combined with computational requirements that exceed practical deployment constraints, motivates the development of efficient vocal-conditioned generation approaches.

Recent developments include OpenAI's Jukebox \cite{dhariwal2020jukebox}, which generates raw audio with artist and genre conditioning but requires over 6 hours for generation \cite{carter2020jukebox}, and Meta's MusicGen \cite{copet2023musicgen}, which provides text and audio conditioning but demands substantial computational resources. Google's SingSong \cite{donahue2023singsong} represents the most relevant prior work, generating instrumental accompaniments from vocal inputs using an adapted AudioLM architecture, but employs autoregressive generation without sophisticated attention mechanisms for capturing complex vocal-accompaniment relationships.

This paper introduces SAMUeL (Soft Alignment Music U-net Latent diffusion), an ultra-lightweight latent diffusion model specifically designed for vocal-conditioned accompaniment generation. The model name reflects its core architectural components: the Soft Alignment attention mechanism that dynamically balances temporal dependencies, the U-net backbone architecture, and operation in compressed latent space for efficiency. SAMUeL directly accepts vocal melodies as input and generates corresponding instrumental backing tracks while maintaining harmonic consistency, rhythmic alignment, and stylistic coherence. SAMUeL operates in compressed latent space using a novel soft alignment attention mechanism that dynamically balances local acoustic patterns with global musical structure.

Our key contributions include: (1) a soft alignment attention mechanism that adaptively combines local and global temporal dependencies based on diffusion timesteps, (2) an end-to-end vocal-conditioned framework bypassing text prompts or symbolic preprocessing, (3) ultra-lightweight architecture achieving 220× parameter reduction with 52× faster inference compared to competing systems, and (4) comprehensive evaluation demonstrating competitive performance despite resource constraints.

\section{Related Work}

\subsection{Diffusion Models for Audio Generation}

Diffusion models have emerged as the dominant paradigm for high-quality audio synthesis, despite computational challenges posed by high-dimensional audio signals and temporal dependencies in musical structure. Early works including DiffWave \cite{kong2020diffwave} and WaveGrad \cite{chen2020wavegrad} demonstrated direct waveform diffusion feasibility but required substantial computational resources.

The development of latent diffusion approaches \cite{rombach2022high} addressed computational limitations by performing denoising in compressed representations learned by variational autoencoders. This approach proves particularly effective for audio applications where perceptual importance varies across frequency components, enabling VAEs to focus on perceptually relevant features.

Recent audio diffusion advances include AudioLM \cite{borsos2023audiolm} introducing discrete token representations, MusicLM \cite{agostinelli2023musiclm} demonstrating large-scale text-conditioned generation, and Stable Audio Open \cite{evans2024stable} providing a comprehensive framework with 156M parameter autoencoder, 109M parameter T5 encoder, and 1057M parameter diffusion transformer. These systems establish diffusion model viability for complex musical tasks but primarily focus on text-based conditioning, limiting applicability for direct musical input.

\subsection{Conditional Audio Generation and Attention Mechanisms}

Text-conditioned approaches \cite{agostinelli2023musiclm, evans2024stable} offer intuitive interfaces but suffer from semantic gaps between textual descriptions and musical concepts. Alternative approaches include chord-based generation \cite{dong2018musegan} providing direct musical control but limited to symbolic representations, and audio-to-audio conditioning focusing on style transfer rather than accompaniment generation \cite{vyas2023audiobox}.

\textbf{Transformer vs. Ours} SAMUeL employs a U-Net architecture augmented with attention mechanisms inspired by transformers \cite{vaswani2017attention} but is distinct in its design. Unlike standard transformers, which rely heavily on Multi-Layer Perceptrons (MLPs) and Layer Normalization contributing significantly to parameter count, SAMUeL omits these components to achieve an ultra-lightweight architecture. Our soft alignment attention mechanism, combining local and global patterns, is tailored for audio applications, leveraging rotary position embeddings (RoPE) \cite{su2024roformer} for global attention while maintaining computational efficiency.

\section{Methodology}

\subsection{Problem Formulation}

We formulate vocal-conditioned accompaniment generation as a conditional diffusion process in latent space. Given a vocal melody $v \in \mathbb{R}^{2 \times T}$ with temporal frames $T$ and 2 channels (stereo audio), we generate a corresponding accompaniment $a \in \mathbb{R}^{2 \times T}$ maintaining harmonic consistency, rhythmic alignment, and stylistic coherence.

The process operates in latent space using a pre-trained music variational autoencoder (VAE), mapping audio to compressed representations $z_v, z_a \in \mathbb{R}^{64 \times T/r}$ with $r$ being the compression rate of the VAE. This enables tractable diffusion modeling while maintaining musical structure necessary for coherent generation.

\subsection{Latent Space Encoding}

We employ the pre-trained VAE component from Stable Audio Open \cite{evans2024stable}, representing state-of-the-art audio compression optimized for latent diffusion generation. The encoder architecture builds on the Descript Audio Codec framework\cite{audiocodec2023}, featuring fully-convolutional design enabling arbitrary-length processing while maintaining high reconstruction fidelity.

The encoder bottleneck is parameterized as a variational autoencoder with 64-dimensional latent space and a compression rate, $r = 2048$, providing continuous representation more suitable for diffusion modeling than discrete tokenization. The decoder employs transposed strided convolutions with symmetric channel contraction and weight normalization throughout. Critically, the decoder omits traditional $\tanh$ activation at output, as empirical evaluation revealed this nonlinearity introduced harmonic distortion artifacts detrimental to musical content quality.

\subsection{Data Collection}
For this study, we introduce POPDB \cite{zhang2025popdb}, a dataset of 15,713 vocal-accompaniment pairs derived from popular songs spanning 2000 onward, totaling approximately 200 hours of audio. The dataset construction leveraged the Song Popularity Dataset \cite{h2021song} from Kaggle, which catalogs popular songs with metadata, selected for its diversity across genres and eras.

Audio was acquired using the yt-dlp Python library for automated YouTube downloading, searching by song titles and filtering out playlists and content exceeding 15 minutes. Audio was processed in WAV format at 44.1 kHz to maintain fidelity. Source separation used Demucs \cite{defossez2020real} to isolate vocal and accompaniment components. Each song was encoded using the Stable Audio Open VAE, producing 1024-token sequences corresponding to 47.55 seconds of audio, then chunked sequentially without overlap. Trailing chunks shorter than 47.55 seconds were discarded.

The 47.55-second chunk size aligns with Stable Audio Open's VAE optimization, balancing efficiency and performance. We hypothesize that at least half of each chunk contains vocals, based on typical song structures where non-vocal segments (e.g., intros, bridges) are usually under 20 seconds \cite{introlengthreport}. This assumption requires validation in future work to ensure robust vocal conditioning. The pipeline required approximately 50 hours of computational time, resulting in POPDB containing 15,713 encoded vocal-accompaniment pairs for training.

\section{Model Architecture}

\subsection{U-Net Design}

SAMUeL employs a U-Net architecture enhanced with novel attention mechanisms specifically designed for vocal-accompaniment relationships. The network processes latent representations $\left(B, 64, 1024\right)$ through a hierarchical encoder-decoder structure featuring multiple resolution levels, residual connections, skip connections, and sophisticated conditioning mechanisms emphasizing dual-stream vocal conditioning.

The U-Net comprises downsampling blocks, a bottleneck attention layer, and upsampling blocks with residual connections. Each downsampling block includes Group Normalization, SiLU activation, Feature-wise Linear Modulation (FiLM)~\cite{perez2018film}, and Conv1D layers, progressively reducing temporal resolution while increasing channel depth: 
\[
(64, 1024) \rightarrow (128, 512) \rightarrow (256, 256) \rightarrow (512, 128).
\]
SiLU (Sigmoid Linear Unit)~\cite{ramachandran2017swish}, also known as Swish, is chosen for its smooth, non-monotonic activation properties, which improve gradient flow and yield better empirical performance in deep sequence models.

This architectural design allows the network to efficiently capture both fine-grained local dependencies and long-range structure in temporal data. By leveraging FiLM and attention-based soft alignment, SAMUeL dynamically integrates conditioning information at multiple scales, enabling more expressive modeling of vocal-guided accompaniment generation.

\subsection{Soft Alignment Attention Mechanism}

The soft alignment attention mechanism represents our core innovation, addressing the challenge of capturing both fine-grained acoustic patterns and global musical structure. Traditional attention mechanisms struggle with extreme sequence lengths in audio applications, while purely local approaches fail to capture long-range musical dependencies such as harmonic progressions and structural relationships.

Our soft alignment attention combines local and global attention patterns through time-dependent weighting reflecting the hierarchical diffusion process. Local attention operates within sliding windows of size 16 on the temporal axis; every chunk is padded with zeros at the beginning and the end. While local attention does not leverage positional embedding to focus on the musical harmonic structure where absolute position is less important, global attention computes relationships across entire sequences using rotary position embeddings \cite{su2024roformer} for improved positional encoding.

The underlying intuition is to expose the model to the full structure of the song early on, while progressively shifting focus toward fine-grained acoustic interactions between nearby temporal frames as diffusion proceeds. This dual-scale attention strategy allows SAMUeL to model both the precise timing details and the broader musical form, which are critical for generating coherent and expressive musical accompaniments.

Local attention for position $i$ is computed as:
\begin{equation}
\text{LocalAttn}(i) = \sum_{j \in W(i)} \text{softmax}\left(\frac{Q_i \cdot K_j}{\sqrt{d}}\right) V_j
\end{equation}
where $W(i)$ represents the local window around position $i$, and $Q$, $K$, $V$ derive from convolutional projections. Global attention incorporates rotary position embeddings:
\begin{equation}
\text{GlobalAttn} = \text{softmax}\left(\frac{\text{RoPE}(Q) \cdot \text{RoPE}(K)^T}{\sqrt{d}}\right) V
\end{equation}

The temporal mixing strategy dynamically combines contexts based on diffusion timestep:
\begin{equation}
\text{Context} = \sqrt{\bar{\alpha}_t} \cdot \text{GlobalAttn} + (1-\sqrt{\bar{\alpha}_t}) \cdot \text{LocalAttn}
\end{equation}
where $\bar{\alpha}_t$ represents the cumulative noise schedule parameter. This ensures early diffusion stages (large $t$) emphasize global attention for overall musical structure, while later stages (small $t$) emphasize local attention for fine-grained detail generation.

\begin{figure}[h]
\centering
\includegraphics[width=0.45\textwidth]{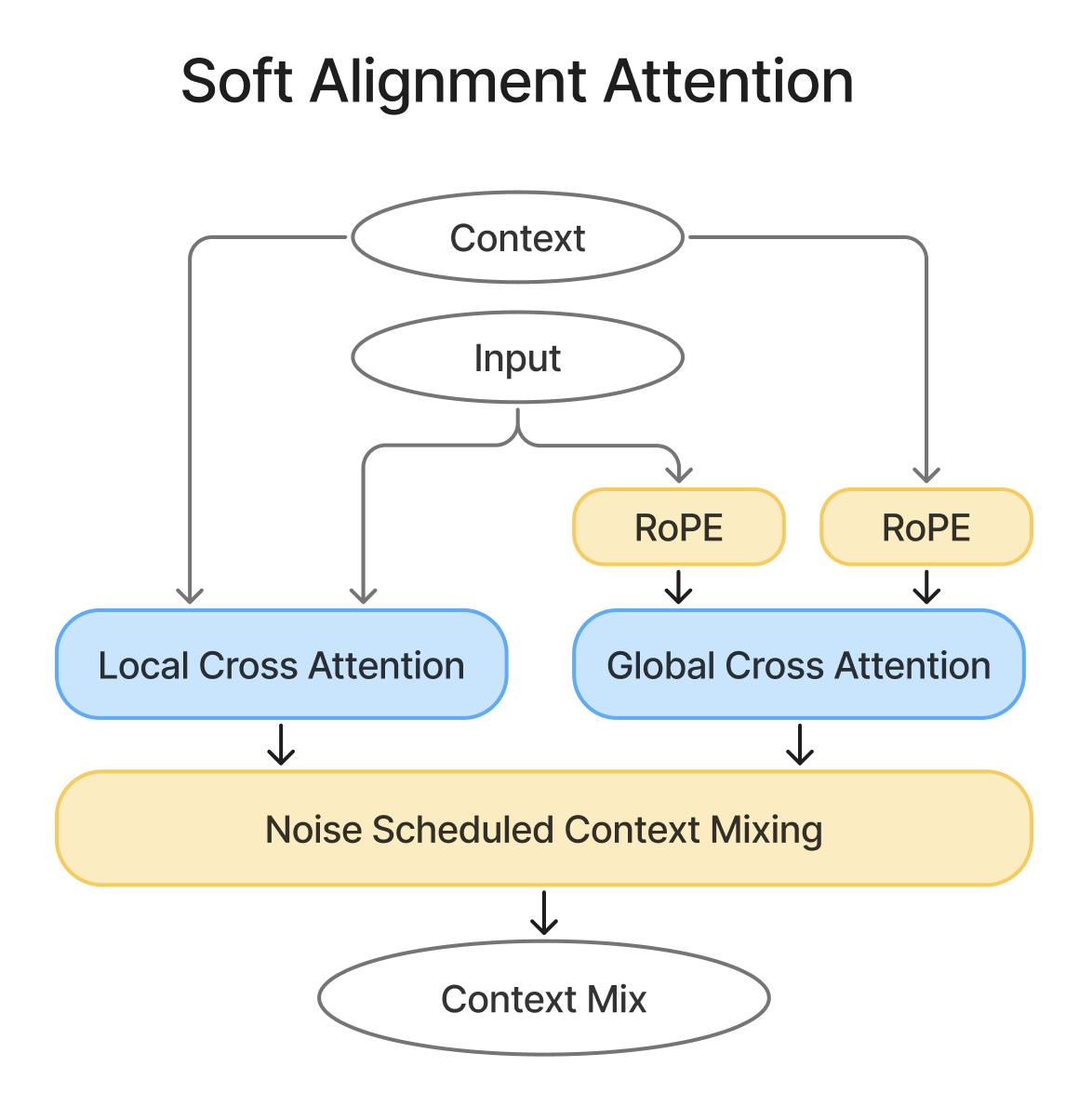}
\caption{Detailed soft alignment attention mechanism computing both local and global attention patterns. The dual-scale approach captures fine-grained temporal dependencies within local windows while maintaining global musical coherence through full-sequence attention, critical for maintaining musical structure and quality. Note that RoPE is not applied to for local attention, learning about the general local acoustic relationships.}
\label{fig:soft_alignment_detail}
\end{figure}

The multi-scale attention design enables SAMUeL to capture information at multiple temporal scales essential for musical coherence and quality. Local attention windows preserve fine-grained acoustic relationships between neighboring temporal frames, while global attention maintains awareness of long-range harmonic progressions and structural elements. This dual-scale approach proves particularly effective for musical applications where both micro-timing relationships and macro-structural coherence are critical for perceptual quality.

\begin{figure}[h]
\centering
\includegraphics[width=0.48\textwidth]{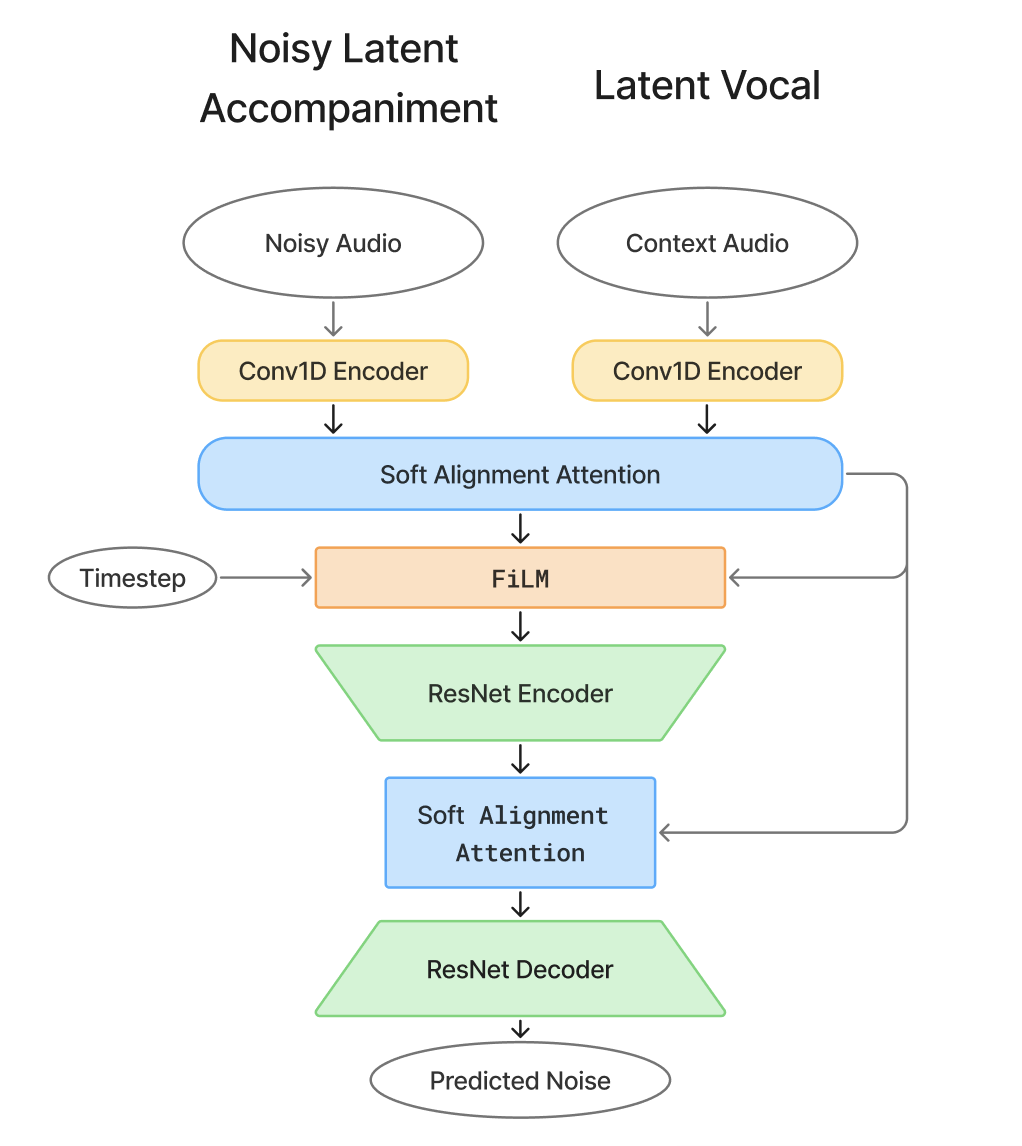}
\caption{Complete model architecture showing U-Net backbone with multi-scale attention mechanisms. Vocal conditioning integrates through cross-attention at multiple stages, while FiLM layers provide fine-grained temporal control.}
\label{fig:model_architecture}
\end{figure}

\subsection{Architecture Specifications and Training Objective}

SAMUeL comprises approximately 15 million parameters distributed across U-Net backbone, attention mechanisms, and conditioning networks. The U-Net backbone contains 3 downsampling and 3 upsampling convolutional blocks, each having two convolutional layers and one FiLM layer. Attention mechanisms utilize 8 heads with 64-dimensional projections, and local attention window size 16 provides effective local modeling while maintaining computational tractability.

SAMUeL employs v-parametrization \cite{salimans2022progressive} rather than standard noise prediction, providing improved training stability and sample quality for audio applications:
\begin{equation}
v_t = \sqrt{\bar{\alpha}_t} \epsilon - \sqrt{1-\bar{\alpha}_t} x_0
\end{equation}
where $\epsilon$ represents noise and $x_0$ represents clean signal. The training objective becomes:
\begin{equation}
\mathcal{L} = \mathbb{E}_{x_0, c, \epsilon, t} \left[ \| v_t - v_\theta(x_t, t, c) \|_2^2 \right]
\end{equation}
where $c$ represents vocal conditioning and $v_\theta$ denotes model prediction.

\section{Training and Implementation}

SAMUeL training employed AdamW optimization \cite{adamw2017} with learning rate $3.5 \times 10^{-4}$ and cosine annealing schedule for one hour on one A100. Batch size selection proved critical for training stability—systematic experimentation revealed batch size 16 provided optimal balance between stable gradient estimation and computational efficiency. This stems from interactions between group normalization layers and signal noise ratio weighted loss function during diffusion.

The training objective employs signal noise weighted MSE loss, dynamically adjusting weighting based on signal-to-noise ratio at each timestep:
\begin{equation}
\mathcal{L}_{\text{SNW}} = \mathbb{E}_{x_0, c, \epsilon, t} \left[ w(t) \| v_t - v_\theta(x_t, t, c) \|_2^2 \right]
\end{equation}
where $w(t) = \frac{\bar{\alpha}_t}{1-\bar{\alpha}_t}$ represents signal-to-noise weighting. This scheme ensures early timesteps focus on global structure learning while later timesteps emphasize fine-grained detail refinement.

\begin{figure}[h]
\centering
\includegraphics[width=0.45\textwidth]{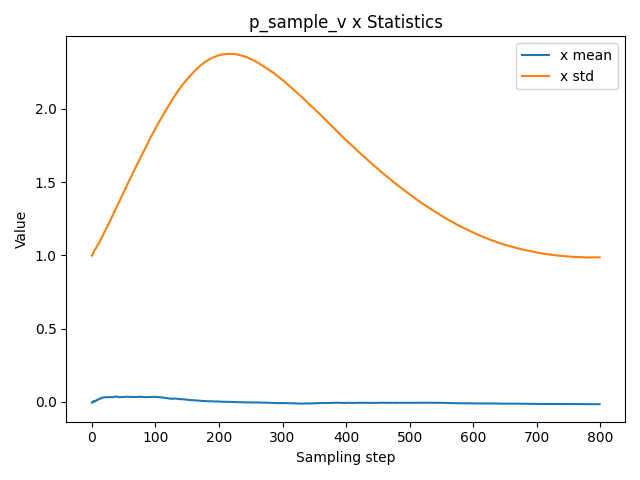}
\caption{Standard deviation of sample prediction $v_\theta$ across diffusion timesteps. High variance during early timesteps reflects global structure learning, while convergence during later timesteps indicates fine-detail refinement.}
\label{fig:sample_stats}
\end{figure}

The diffusion process utilized cosine noise schedule with 800 timesteps, and classifier-free guidance training incorporated 10\% conditioning dropout probability. Figure \ref{fig:sample_stats} revealed the reconstructed audio distribution returns to a normal distribution, $\cal{N} (0, I)$, at around 600 timesteps; an 800 diffusion timesteps is chosen to completely reconstruct the latent distribution most efficiently. A 10\% held-out validation set monitored training progress with early stopping based on validation loss convergence. Mixed precision training provided computational efficiency while maintaining numerical stability.

\section{Experimental Results}

\subsection{Evaluation Methodology and Quantitative Results}

We evaluated generated accompaniments using AudioBox-Aesthetics \cite{vyas2023audiobox}, a pre-trained transformer-based model developed by Meta that provides comprehensive audio quality assessment. This model addresses critical limitations of existing audio evaluation metrics such as Fréchet Audio Distance (FAD), which produces only overall scores without capturing the nuanced aspects of musical quality. Unlike traditional benchmarks that provide uninformative evaluations and amplify individual bias from human-annotated data, AudioBox-Aesthetics represents the only available open-source pre-trained model capable of multi-dimensional audio evaluation.

The AudioBox-Aesthetics model provides objective metrics across four independent dimensions, each scored on a scale from 1 (less desirable) to 10 (more desirable): Production Quality (PQ) measuring technical fidelity and recording quality, Production Complexity (PC) assessing arrangement sophistication and musical intricacy, Content Enjoyment (CE) evaluating subjective appeal and listener satisfaction, and Content Unity (CU) measuring musical coherence and structural consistency. The model was trained on 500 hours of audio samples annotated by 158 expert raters, ensuring robust evaluation standards. The training process emphasized category independence, with low correlation between evaluation dimensions demonstrating that each metric captures distinct aspects of audio quality. Rating distributions span the full evaluation scale, indicating diversity in the collected training data and comprehensive coverage of quality variations.

The original validation study confirmed high correlation between the model's predicted quality scores and ground truth labels from expert human evaluators, establishing its reliability for objective music quality assessment. This multi-dimensional approach proves particularly valuable for evaluating generative music models, where different aspects of quality may vary independently based on architectural choices and training constraints.

Evaluation compared SAMUeL against Stable Audio Open, MusicGen, OpenAI Jukebox, and ground truth across 10 generated samples, and the average score of the ten tracks are reported in Table \ref{tab:quality_metrics}. Due to the small sample size, results are preliminary and require larger-scale validation in future work.

\begin{table}[h]
\centering
\caption{Music Quality Assessment and Model Comparison}
\begin{tabular}{@{}lcccc@{}}
\toprule
Model/Source & PQ & PC & CE & CU \\
\midrule
Pop Benchmark & 7.92 & 6.26 & 7.36 & 7.61 \\
MusicGen & 7.99 & 5.62 & 7.99 & 8.14 \\
Stable Audio Open & 7.54 & 4.06 & 6.84 & 7.71 \\
OpenAI Jukebox & 5.30 & 5.97 & 4.83 & 4.66 \\
SAMUeL & 5.73 & 5.14 & 2.93 & 4.97 \\
\bottomrule
\end{tabular}
\label{tab:quality_metrics}
\end{table}
\begin{table}[h]
\centering

\begin{threeparttable}
\caption{Efficiency and Parameter Analysis}
\begin{tabular}{@{}lccc@{}}
\toprule
Model & Parameters & Inference Time & Conditioning \\
\midrule
OpenAI Jukebox & 5.0B & $>$6h\tnote{1} & Lyrics + Genre \\
Stable Audio Open & 1.2B & 25s & Text Description \\
MusicGen - Large & 3.3B & 1083s & Vocal + Text \\
MusicGen - Small & 300M & 236s & Vocal + Text \\
SAMUeL & 15M & 21s & Vocal Melody \\
\bottomrule
\end{tabular}
\begin{tablenotes}
\small
\item[1] Result taken from \cite{carter2020jukebox}
\end{tablenotes}
\label{tab:model_comparison}
\end{threeparttable}
\end{table}

Results reveal competitive performance considering SAMUeL's smaller scale and training constraints. While achieving lower absolute scores compared to larger systems, SAMUeL outperforms OpenAI Jukebox in Production Quality (5.73 vs 5.30) and Content Unity (4.97 vs 4.66), attributed to Jukebox's tendency toward noise artifacts. The Production Complexity score (5.14) demonstrates that despite resource constraints, soft alignment attention enables reasonably sophisticated musical arrangements through direct vocal conditioning.

SAMUeL achieves exceptional efficiency with only 15M parameters—220× smaller than MusicGen Large, 20× smaller than MusicGen Small, 80× smaller than Stable Audio Open, and 333× smaller than OpenAI Jukebox. Inference speed of 21 seconds per 47-second audio represents substantial advantages: 52× faster than MusicGen Large and 11× faster than MusicGen Small. While Stable Audio Open achieves comparable inference time, it cannot process direct musical input, positioning SAMUeL as optimal for audio-to-audio conditioning.

\subsection{Qualitative Analysis and Performance Trade-offs}

Subjective evaluation reveals effective vocal-accompaniment alignment regarding harmonic progression consistency and rhythmic synchronization. The soft alignment attention mechanism successfully establishes appropriate temporal relationships between vocal phrases and accompaniment responses, while global context modeling maintains structural coherence throughout generation.

Generated accompaniments exhibit characteristic properties reflecting both strengths and limitations. Audio outputs predominantly feature synthesizer-like timbres, suggesting effective electronic texture generation within latent space constraints. Percussion elements are consistently present and rhythmically aligned with vocal input, though exhibiting noisy characteristics rather than clean, distinct drum sounds.

A notable limitation is SAMUeL's current inability to generate distinct instrumental sounds with clear acoustic characteristics, reflected in lower content enjoyment scores compared to models trained on larger datasets. Generated accompaniments lack timbral diversity typical of professional recordings, producing more homogeneous sonic palettes dominated by synthesized textures. This limitation stems from compressed latent representation, limited training data, and focus on overall musical structure rather than fine-grained timbral detail—constraints inherent to ultra-lightweight design philosophy.

Despite these constraints, vocal conditioning proves effective for establishing foundational musical relationships. Direct musical input successfully bypasses semantic gaps inherent in text-based approaches, enabling precise control over harmonic progressions and rhythmic alignment difficult to achieve through natural language descriptions.

\section{Discussion and Future Work}

SAMUeL demonstrates direct musical input viability for accompaniment generation, addressing a significant gap in current music AI systems. The soft alignment attention mechanism provides effective solutions for capturing multi-scale temporal dependencies in audio, while v-parametrization ensures stable training dynamics suitable for complex musical content.

\subsection{Efficiency Advantages and Deployment Considerations}

A key distinguishing feature is exceptional efficiency compared to existing systems. The ultra-lightweight architecture enables deployment on consumer-grade hardware with minimal memory requirements (2-4GB GPU memory), making professional-quality musical accompaniment generation accessible for mobile devices, edge computing, and resource-constrained environments. This democratization represents significant advancement over existing models requiring enterprise-level computational resources.

Performance trade-offs inherent in efficiency gains are evident in AudioBox-Aesthetics evaluation, where SAMUeL achieves lower absolute scores compared to larger systems. However, the ability to outperform OpenAI Jukebox demonstrates that architectural innovation can compensate for scale limitations. Relatively modest scores in content enjoyment and unity reflect challenges of training with fewer parameters and limited data, yet SAMUeL successfully synthesizes complete multi-track accompaniments—remarkable given resource constraints.

\subsection{Training Insights and Limitations}

Experimentation revealed important phenomena regarding model architecture and training dynamics. A critical finding concerns the relationship between model complexity and perceptual quality: increasing U-Net depth consistently marginally reduced training loss but did not correlate with improved audio quality. This suggests traditional loss metrics may inadequately capture perceptual aspects of musical generation, highlighting fundamental challenges where mathematical optimization objectives diverge from human preferences. Further works could explore alternative training objectives that better align with perceptual quality, such as adversarial training or perceptual loss functions.

Current limitations include fixed-length segment operation limiting applicability to longer compositions requiring structural coherence across sections. The 47-second chunk constraint limits capture of extended musical narratives and relationships spanning longer time scales. Additionally, vocal conditioning could be enhanced through integration with lyrics, chord progressions, or explicit style controls.

A significant constraint affecting SAMUeL's performance stems from limited training data due to computational resource restrictions. Our model was trained on approximately 200 hours of audio data, substantially less than the 7,300 hours utilized by Stable Audio Open \cite{evans2024stable}. This 36-fold disparity in training data represents a critical limiting factor, as deep generative models typically exhibit strong scaling behavior with increased data availability. Despite this constraint, SAMUeL demonstrates competitive performance relative to its training data scale, suggesting that the sophisticated soft alignment attention architecture effectively leverages available information. We anticipate that scaling to comparable dataset sizes would yield substantial performance improvements, particularly in timbral diversity and content enjoyment metrics where data-driven learning is most beneficial.

Future work should explore hierarchical generation approaches handling multiple temporal scales simultaneously, enabling coherent composition of complete songs. Variable-length audio processing could increase training data availability and reduce computational requirements. Multi-modal conditioning combining vocal, textual, and symbolic inputs could provide enhanced control for professional workflows.

\section{Conclusion}

We have presented SAMUeL (Soft Alignment Music U-net Latent diffusion), a novel latent diffusion model for vocal-conditioned musical accompaniment generation addressing key limitations in existing music AI systems. The soft alignment attention mechanism effectively captures both local acoustic patterns and global musical structure through adaptive temporal mixing, while v-parametrization provides stable training dynamics. Operating in compressed latent space achieves computational efficiency suitable for interactive applications while maintaining audio quality through sophisticated conditioning mechanisms.

SAMUeL's ultra-lightweight 15M parameter architecture demonstrates significant efficiency advantages: 52× faster inference than MusicGen Large while using 220× fewer parameters, and 11× faster than MusicGen Small while using 20× fewer parameters. While absolute perceptual quality scores are lower than large-scale models due to training constraints, SAMUeL outperforms OpenAI Jukebox and achieves reasonable performance considering exceptional efficiency. This makes it uniquely suitable for real-time creative workflows, mobile applications, and resource-constrained environments where existing models are computationally prohibitive.

This work establishes vocal conditioning as a viable paradigm for efficient music generation and provides foundation for future developments in AI-assisted composition and interactive music systems prioritizing accessibility and real-time capability. SAMUeL's design philosophy—combining architectural innovation with computational efficiency—opens new possibilities for democratizing AI-powered music creation.

\section{Acknowledgment}

The authors would like to acknowledge the funding and support from Mitacs and the University of Toronto Engineering Science Division. We would like to the support from Arthur Chan, our home supervisor for organizing our research opportunity, the Engineering Science Research Opportunity Program - Global at the King Mongkut's University Technology Thonburi.



\begin{thebibliography}{99}

\bibitem{agostinelli2023musiclm} A. Agostinelli, T. I. Denk, Z. Borsos, J. Engel, M. Verzetti, A. Caillon, Q. Huang, A. Jansen, A. Roberts, M. Tagliasacchi, M. Sharifi, N. Zeghidour, and C. Frank, ``MusicLM: Generating Music From Text,'' in \textit{International Society for Music Information Retrieval Conference}, 2023.

\bibitem{copet2023musicgen} J. Copet, F. Kreuk, I. Gat, T. Remez, D. Kant, G. Synnaeve, Y. Adi, and A. Défossez, ``Simple and Controllable Music Generation,'' in \textit{Advances in Neural Information Processing Systems}, vol. 36, 2023.

\bibitem{dong2018musegan} H.-W. Dong, W.-Y. Hsiao, L.-C. Yang, and Y.-H. Yang, ``MuseGAN: Multi-track Sequential Generative Adversarial Networks for Symbolic Music Generation and Accompaniment,'' in \textit{Proc. AAAI Conference on Artificial Intelligence}, vol. 32, 2018.

\bibitem{evans2024stable} Z. Evans, C. J. Carr, J. Taylor, S. H. Hawley, and J. Pons, ``Stable Audio Open,'' in \textit{International Conference on Machine Learning}, 2024.

\bibitem{borsos2023audiolm} Z. Borsos, R. Marinier, D. Vincent, E. Kharitonov, O. Pietquin, M. Sharifi, D. Roblek, O. Teboul, D. Grangier, M. Tagliasacchi, and N. Zeghidour, ``AudioLM: A Language Modeling Approach to Audio Generation,'' \textit{IEEE/ACM Trans. Audio, Speech, and Language Processing}, vol. 31, pp. 2523--2533, 2023.

\bibitem{dhariwal2020jukebox} P. Dhariwal, H. Jun, C. Payne, J. W. Kim, A. Radford, and I. Sutskever, ``Jukebox: A Generative Model for Music,'' in \textit{International Conference on Machine Learning}, 2020.

\bibitem{carter2020jukebox} R. B. Carter, ``A Simple OpenAI Jukebox Tutorial for Non-Engineers,'' 2020. [Online]. Available: \url{https://robertbrucecarter.com/writing/2020/07/a-simple-openai-jukebox-tutorial-for-non-engineers/}

\bibitem{donahue2023singsong} C. Donahue, I. Simon, and S. Dieleman, ``SingSong: Generating Musical Accompaniments From Singing,'' \textit{arXiv preprint arXiv:2301.12662}, 2023.

\bibitem{kong2020diffwave} Z. Kong, W. Ping, J. Huang, K. Zhao, and B. Catanzaro, ``DiffWave: A Versatile Diffusion Model for Audio Synthesis,'' in \textit{International Conference on Learning Representations}, 2021.

\bibitem{chen2020wavegrad} N. Chen, Y. Zhang, H. Zen, R. J. Weiss, M. Norouzi, and W. Chan, ``WaveGrad: Estimating Gradients for Waveform Generation,'' in \textit{International Conference on Learning Representations}, 2021.

\bibitem{rombach2022high} R. Rombach, A. Blattmann, D. Lorenz, P. Esser, and B. Ommer, ``High-Resolution Image Synthesis with Latent Diffusion Models,'' in \textit{Proc. IEEE/CVF Conference on Computer Vision and Pattern Recognition}, pp. 10684--10695, 2022.

\bibitem{vyas2023audiobox} A. Tjandra, S. Sakti, and S. Nakamura, ``Meta Audiobox Aesthetics: Unified Automatic Quality Assessment for Speech, Music, and Sound,'' \textit{arXiv preprint arXiv:2502.05139}, 2025.

\bibitem{vaswani2017attention} A. Vaswani, N. Shazeer, N. Parmar, J. Uszkoreit, L. Jones, A. N. Gomez, L. Kaiser, and I. Polosukhin, ``Attention Is All You Need,'' in \textit{Advances in Neural Information Processing Systems}, vol. 30, 2017.

\bibitem{su2024roformer} J. Su, Y. Lu, S. Pan, A. Murtadha, B. Wen, and Y. Liu, ``RoFormer: Enhanced Transformer with Rotary Position Embedding,'' \textit{Neurocomputing}, vol. 568, pp. 127063, 2024.

\bibitem{audiocodec2023} A. Défossez, J. Copet, G. Synnaeve, and Y. Adi, ``High Fidelity Neural Audio Compression,'' in \textit{International Conference on Machine Learning}, 2022.

\bibitem{zhang2025popdb} B. Zhang, ``POPDB: Encoded Songs Dataset,'' Kaggle, 2024. [Online]. Available: \url{https://www.kaggle.com/datasets/boyazhangnb/encodedsongs}

\bibitem{h2021song} H. Yasser, ``Song Popularity Dataset,'' Kaggle, 2021. [Online]. Available: \url{https://www.kaggle.com/datasets/yasserh/song-popularity-dataset}

\bibitem{defossez2020real} A. Défossez, G. Synnaeve, and Y. Adi, ``Real Time Speech Enhancement in the Waveform Domain,'' in \textit{Proc. Interspeech}, 2020.

\bibitem{introlengthreport} Hit Songs Deconstructed, ``Hit Songs Deconstructed Trend Report Q2 2015,'' 2015. [Online]. Available: \url{https://reports.hitsongsdeconstructed.com/2015/07/30/hit-songs-deconstructed-trend-report-q2-2015/\#averageintrolength}

\bibitem{perez2018film} E. Perez, F. Strub, H. de Vries, V. Dumoulin, and A. Courville, ``FiLM: Visual Reasoning with a General Conditioning Layer,'' in \textit{Proc. AAAI Conference on Artificial Intelligence}, vol. 32, 2018.

\bibitem{ramachandran2017swish} P. Ramachandran, B. Zoph, and Q. V. Le, ``Searching for activation functions,'' \textit{arXiv preprint arXiv:1710.05941}, 2017.

\bibitem{salimans2022progressive} T. Salimans and J. Ho, ``Progressive Distillation for Fast Sampling of Diffusion Models,'' in \textit{International Conference on Learning Representations}, 2022.

\bibitem{adamw2017} I. Loshchilov and F. Hutter, ``Decoupled Weight Decay Regularization,'' in \textit{International Conference on Learning Representations}, 2019.

\end{thebibliography}
\end{document}